# Magnetic structure of quasi-1D antiferromagnetic TaFe$_{1+y}$Te$_3$ with two-leg zigzag ladders

X. Ke[1], B. Qian[2], H. Cao[1], J. Hu[2], G.C. Wang[2], and Z. Q. Mao[2]

[1]*Quantum Condensed Matter Division, Oak Ridge National Laboratory, Oak Ridge, TN 37831, USA*

[2]*Department of Physics and Engineering Physics, Tulane University, New Orleans, Louisiana 70118, USA*

We report the magnetic structure of TaFe$_{1+y}$Te$_3$ single crystals by means of neutron diffraction measurements. TaFe$_{1+y}$Te$_3$ possesses a layered structure with a formation of two-leg zigzag ladders along the *b*-axis. We find that TaFe$_{1+y}$Te$_3$ undergoes an antiferromagnetic transition at 178 K with Fe1 spins of the intra-ladders ferromagnetically aligned while spins of the inter-ladders antiferromagneitcally coupled. Furthermore, spins of the neighboring interstitial Fe2 (y) ions order parallel to the Fe1 spins of each ladder. These findings are distinct from the magnetic structure of the recently-discovered spin-ladder compound BaFe$_2$Se$_3$. TaFe$_{1+y}$Te$_3$ may serve as an interesting quasi-one dimensional *ferromagnetic* system.

There has been intense interest in searching for new iron-based superconductors since the initial discovery of superconductivity in La($O_{1-x}F_x$)FeAs in 2008 [1]. Many types of iron pnictide and iron chalcogenide superconductors have been discovered, including LnFeAs(O,F) (Ln = lanthanide) (1111) [2], (A,K/Na)$Fe_2As_2$ (A = Ba, Sr, Ca, Eu) [3] and (Ba/Sr/Ca)(Fe,TM)$_2As_2$ (TM = Co, Ni, Rh, Pd, Ir, Ru, and Pt) (122) [4,5], $A_{1-x}$FeAs (A = Li or Na) (111) [6], and $Sr_2VO_3FeAs$ [7], and $Fe_{1+y}$(Te, Se) (11) [8]. These materials share a common structural characteristic: Fe tetrahedrally coordinated by As or (Te, Se) to form square-planar sheets. The consensus is that magnetism and superconductivity are intimately correlated and compete with each other in these materials, as evidenced by the enhanced spin fluctuation above $T_c$ [9,10,11,12,13,14,15,16,17], the emergence of spin resonance [11,13,14,15] and the suppression of magnetism below $T_c$ [18]. Studies of the magnetic structure and spin dynamics of these materials have played a key role in understanding mechanisms of superconductivity and exploring for new superconductors.

In addition to 1111-, 122-, 111- and 11-type materials, several other types of Fe-based materials have recently been investigated, including $A_2Fe_4Se_5$ [19,20,21,22,23] with A = Rb, Cs, (Tl, Rb / K), $BaFe_2Se_3$ [24,25,26], $TaFe_{1+y}Te_3$ [27]. $A_2Fe_4Se_5$ has a tetragonal structure, with Fe vacancies forming a $\sqrt{5}\times\sqrt{5}\times1$ supercell structure [21,22]; it exhibits an antiferromagnetic (AFM) transition at $T_N$ = 560 K, followed by a superconducting transition at 30K. The AFM state of this material is characterized by a checkerboard-like magnetic structure formed of ferromagnetically aligned four-spin blocks that are antiferromagnetically coupled to neighboring blocks [21,22]. The crystal structure of $BaFe_2Se_3$ [24,25,26] is also distinct from those of 1111-, 122-, 111- and 11-type materials; it is composed of double chains of $FeSe_4$ edge-sharing tetrahedra and can be regarded as a quasi-one dimensional system. $BaFe_2Se_3$ undergoes a paramagnet-antiferromagnet phase transition with $T_N \sim 240$ K [24,25], where four Fe spins within each two-leg ladder form a checkerboard-like block and align parallel to each other within each block while antiferromagnetically coupled to their nearest neighbors [25], a feature similar to that observed in $A_2Fe_4Se_5$ [21,22]. Nevertheless, the small superconducting feature previously reported by Krzton-Maziopa *et al.* [24] was not observed in later studies [25,26] where it was argued that the superconducting feature in $BaFe_2Se_3$ is an artificial effect and associated with existence of impurities. A more recent study of $BaFe_{2-\delta}Se_3$ single crystals [26] did not find the antiferromagnetic transition, which is presumably attributed to the Fe deficiency.

$TaFe_{1+y}Te_3$, the material studied in this article, was discovered [28] about two decades ago and was recently revisited by Liu *et al.* [27]. This compound possesses a $P2_1/m$ monoclinic crystal structure, with the lattice parameters a = 7.436 Å, b = 3.638 Å, c = 10.008 Å, and $\beta$ = 109.17º [28]. The Ta-Fe bonded network lies between Te layers forming a $FeTaTe_3$ 'sandwich' [27,28], as shown in Fig. 1(a). The excess Fe ($y$) ions partially occupy a square pyramidal site. Similar to $BaFe_2Se_3$ [24,25,26], Fe ions also form two-leg ladders along a principle axis ($b$-axis) in $TaFe_{1+y}Te_3$, but with a zigzag shape instead of rectangular one, thus representing another intriguing quasi-one dimensional magnetic system. $TaFe_{1.25}Te_3$ ($y$ = 0.25) has a structural phase transition at 1010 K, and orders antiferromagnetically below 200 K [28]. Interestingly, this material displays metallic behavior down to 4 K [27,28]. Detailed susceptibility, magnetoresistance, and Hall effect measurements [27] suggest that the AFM transition is of a spin-density-wave character and that the Fe1 moment is about 3.7 $\mu_B$/Fe and the Fe2 ((*i.e.* interstitial Fe ion)) moment is about 4 $\mu_B$/Fe. Furthermore, it was proposed that neighboring spins within each zigzag ladder aligned antiferromagnetically while spins between neighboring ladders are ferromagnetically coupled [27]. However, this needs to be validated by other techniques, such as neutron scattering studies, which are not available yet until this work.

In this article we report the magnetic structure of TaFe$_{1+y}$Te$_3$ ($y$ = 0.17) revealed by single crystal neutron diffraction measurements. In sharp contrast to what has been proposed by Liu *et al.* [27], we find that in the AFM state of TaFe$_{1+y}$Te$_3$, the Fe spins within each ladder are aligned parallel to each other while spins between ladders are antiferromagnetically coupled. Furthermore, the magnetic moment of interstitial Fe2, which are randomly sited, also prefers to be parallel to Fe1 spins of each ladder, as illustrated in Fig. 1(b). This suggests a strong ferromagnetic exchange interaction of Fe1 spins along the zigzag rungs ($J_{nn}$), rendering the system to be a quasi-one dimensional ferromagnet. Such a peculiar magnetic structure is dramatically different from that of BaFe$_2$Se$_3$ with a crystal structure also composed of two-leg ladders.

Single crystals of TaFe$_{1+y}$Te$_3$ were grown using chemical vapor transport method, as described in earlier literature [27,28]. Powders of the raw materials Ta, Fe, and Te with a nominal molar ratio of 1:1.25:3 were ground and then sealed in an evacuated quartz tube together with TeCl$_4$ that serves as transporting agent. The tube was then placed in a furnace and slowly heated up with the hot end at 690 °C and the cool end at 660 °C. The furnace was cooled to room temperature after one week of growth time. The typical dimensions of single crystals grown with this method are ~3 × 4 × 0.5 mm$^3$. The structure of the crystals was characterized by X-ray diffraction. The electronic and magnetic properties of crystals were measured using Quantum Design PPMS and SQUID magnetometer, respectively. To obtain the magnetic structure of this material, a single crystal with a mass of ~ 11 mg was measured using the four-circle neutron diffractometer HB-3A located at the High Flux Isotope Reactor, Oak Ridge National Laboratory.

A neutron wavelength of 1.536 Å [29] was used, unless noted otherwise, by using a double focusing Si(2 2 0) monochromator.

Figure 2(a) shows the temperature dependence of magnetization of TaFe$_{1+y}$Te$_3$ measured with a magnetic field of 1000 Oe applied along in-plane (IP) and out-of-plane (OOP) directions. Note that the OOP direction is about 17.6º degree tilt from the [1 0 -1] direction, as shown in Figure 1. The magnetization shows a maximum around 178 K and field-cool and zero-field-cool measurements do not show any noticeable difference, which indicates the onset of an antiferromagnetic transition. As noted above, the previously-reported AFM transition temperature $T_N$ for a TaFe$_{1.25}$Te$_3$ powder sample is ~ 200K, about 20K higher than the transition temperature observed in our sample; this discrepancy may be due to the lower Fe2 concentration ($y < 0.25$) in our sample, as confirmed by the neutron diffraction measurements shown below. The larger suppression of magnetization with the field along the OOP direction than that along the IP direction suggests the nature of magnetic anisotropy with the spin easy axis tilt towards the OOP direction.

In Fig. 2(b) we plot the resistivity as a function of temperature measured with a DC current (I = 1 mA) applied along the IP and OOP directions. The data were taken using a standard four-probe method. For the current applied along the IP direction, the material exhibits metallic behavior over the whole measured temperature range. In addition, the AFM transition results in a steeper decrease in resistivity and a kink near $T_N$. These characteristics are indicative of an itinerant antiferromagnet. However, the resistivity along the OOP direction of most samples we measured exhibits non-metallic behavior in the whole temperature range (main panel of Fig. 2b), with $\rho_{OOP} / \rho_{IP} \approx 50$ at $T = 2K$. Such an anisotropic behavior in electronic transport is associated with the layered crystal structure and magnetic structure as discussed below.

Occasionally, a metallic feature is observed at low temperature along the OOP direction, as shown in the inset of Fig. 1(b), which may originate from rich excess Fe that helps inter-plane bonding and enhance conductivity.

To characterize the nuclear and magnetic structure of TaFe$_{1+y}$Te$_3$, we have performed single-crystal neutron diffraction measurements at various temperatures between 5 K and room temperature. The crystal structure refined from the neutron scattering data collected at 5 K (Fig. 1(a)) does not show any essential difference from the room temperature structure except for a slight thermal contraction of the lattice. Data refinement using Fullprof [30] with the refinement goodness shown in Fig. 4(a) reveals a smaller concentration of interstitial Fe ions than the expected nominal value, with $y$ = 0.172 (8), which may explain the lower $T_N$ value in our single crystal sample as compared to the previously-reported value (~ 200 K) for TaFe$_{1.25}$Te$_3$ [28]. Furthermore, no superlattice peaks are observed, indicative of the random occupancy of Fe2 interstitials; this is consistent with the previous x-ray and TEM results [28].

In addition to the nuclear Bragg diffraction, neutron scattering intensities also show peaks in (H K L) with half integer values of H and L. For instance, Fig. 3(a) plots the rocking curve measurements of (0.5 0 0.5) and (-0.5 0 0.5) peaks taken at $T$ = 5 K using a neutron wavelength of 2.410 Å to avoid the half λ contamination, which shows nice Gaussian shape with the full width at half maximum defined by the instrumental resolution. Note that the magnetic form factor associated with the magnitude of (0.5 0 0.5) and (-0.5 0 0.5) vectors is almost the same, thus, the difference in the diffraction intensity of these two $Q$ vectors originates from their relative direction to the magnetic moment. Such diffractions with half integer values of H and L are ascribed to the antiferromagnetic magnetic diffractions. This is clearly evidenced by the temperature dependence of (0.5 0 0.5) diffraction intensity shown in Fig. 3(b), and the gradual increase in intensity below $T_N \sim 178$ K is characteristic of a second order phase transition, in agreement with both transport and magnetic susceptibility measurements presented in Fig. 2.

We have measured a series of magnetic diffraction peaks at $T = 5$ K to determine and refine the magnetic structure of TaFe$_{1+y}$Te$_3$. The magnetic ordering propagation vector is determined to be (-0.5 0 0.5), based on which, one can obtain 4 irreducible representations to describe the magnetic structure using the BasIresps program in Fullprof [30]. These include parallel / antiparallel spin alignment along the *b*-axis or the *ac*-plane. We have refined the magnetic diffraction data (including 40 magnetic reflections) in terms of these four possible magnetic structures and find that only the magnetic structure shown in Fig. 1b can yield a good fit to the data, as manifested in the consistency of calculated and measured intensity displayed in Fig. 4b. This magnetic structure possesses the following remarkable characteristics: i) Fe1 spins along the chain direction (*b*-axis) are parallel; ii) Fe1 spins of two neighboring chains also point in a parallel direction, thus forming a *ferromagnetic* two-leg zigzag ladder; iii) spin direction of neighboring interstitial Fe2 of each ladder prefers to align parallel to the Fe1 spin direction; iv) spins of neighboring zigzag ladders align antiparallel to each other in the *ac*-plane. A closer look of the Fe spin configuration is plotted in Fig. 1(c). The magnetic moment points along the [1 0 -1] direction, consistent with the magnetic susceptibility results plotted in Fig. 2(a) that shows a larger magnetic susceptibility value along the OOP direction. And the moment size extracted from the data refinement is 2.1 (1) $\mu_B$ / Fe for Fe1 and 2.6 (1) $\mu_B$ / Fe for Fe2, both of which are smaller than the expected values for the high spin states of Fe$^{2+}$ (3d$^4$) and Fe$^{3+}$ (3d$^5$). Note that the valence values of Fe1 and Fe2 may be a mixture of both Fe$^{2+}$ and Fe$^{3+}$. The suppression of magnetic moment is presumably associated with the itinerancy of charge carriers as evidenced by the metallic electronic transport feature shown in Fig. 2(b).

Such a magnetic structure of TaFe$_{1+y}$Te$_3$ is in sharp contrast to the one proposed recently by Liu *et al* [27] that is composed of antiferromagnetic zigzag chains of Fe1 with the neighboring ladders couple ferromagnetically below $T_N$. It is also distinct from the antiferromagnetically-coupled checkerboards consisting of 4 ferromagnetically-aligned spins observed in BaFe$_2$Se$_3$ [25] which is also a quasi-one dimensional system but with an orthorhombic crystal structure. It suggests that the magnetic coupling of the nearest-neighboring Fe1 spins of the zigzag ladders in TaFe$_{1+y}$Te$_3$, $J_{nn}$ shown in Fig. 1(c), are ferromagnetic, which may be dominated by the direct exchange interaction between Fe1 spins considering the short Fe1-Fe1 distance (2.72 Å) that is slightly longer than the interatomic distance of Fe metal (~ 2.53 Å). In addition, the exchange interaction between the next nearest-neighboring Fe1 spin along the chain direction, $J_{nnn}$, may be ferromagnetic as well mainly due to the almost 90°-exchange path of Fe1-Te-Fe1. We speculate that the parallel spin alignment of Fe2 to Fe1 may originate from the ferromagnetic direct exchange interaction owing to their short distance (~ 2.49 Å). These ferromagnetic exchange interactions lead to the parallel spin alignment of each zigzag ladder and the Fe2 interstitials, which consequently inhibits the occurrence of superconductivity at low temperatures. Detailed first principles calculations and inelastic neutron scattering measurements are warranted to clarify the nature of these magnetic interactions.

Finally, let's turn to discuss the magnetic coupling between neighboring ladders. The antiparallel spin alignment between neighboring ladders indicates an antiferromagnetic interaction that induces the observed paramagnetic-antiferromagnetic transition below $T_N \sim 178$ K. As shown in Fig. 1(a), however, we speculate that the superexchange interaction between Fe1 ions of neighboring ladders along both the out-of-plane direction (interlayer) and the in-plane direction is relatively weak and much smaller than the energy scale of the transition temperature, considering that the shortest distance of these Fe1 ions are ~ 6.78 Å and 8.54 Å, respectively. Thus, TaFe$_{1+y}$Te$_3$ can be regarded as a quasi-one dimensional ferromagnetic system. This appears to be a one-dimensional analog of the quasi-two dimensional ferromagnetic Ca$_3$Ru$_2$O$_7$ [31,32] where ferromagnetic coupled bilayers are stacked antiferromagnetically along the out-of-plane direction. A possible mechanism that drives the antiferromagnetic transition in TaFe$_{1+y}$Te$_3$ is *via* the superexchange interaction involving Fe2 interstitials, which requires further investigations.

In summary, we have measured the magnetic structure of layered TaFe$_{1+y}$Te$_3$ and find that it is composed of ferromagnetic two-leg zigzag ladders that are antiferromagnetically coupled to their neighbors along both in-plane and out-of-plane directions. This contrasts to the scenario proposed recently by Liu *et al.* [27]. TaFe$_{1+y}$Te$_3$ may serve as an interesting quasi-one dimensional ferromagnetic system.

Research at Oak Ridge National Laboratory's High Flux Isotope Reactor was sponsored by the Scientific User Facilities Division, Office of Basic Energy Sciences, U.S. Department of Energy. Work at Tulane is supported by the NSF under grant DMR-0645305 and the LA-SiGMA program under award #EPS-1003897. X. K. gratefully acknowledges the financial support by the Clifford G. Shull Fellowship at ORNL.

FIGURE CAPTIONS

Figure 1. Schematics of the crystal structure (a) and spin structure (b), and detailed view of the zigzag ladders (c) of TaFe$_{1+y}$Te$_3$. Inset in (b) shows the out-of-plane (OOP) direction tilts from the [1 0 -1] direction by about 17.6º degree.

Figure 2. Temperature dependence of magnetization (a) and resistivity (b) of TaFe$_{1+y}$Te$_3$ along both in-plane (IP) and out-of-plane (OOP) directions.

Figure 3. (a) Rocking curve of magnetic reflections (-0.5 0 0.5) and (0.5 0 0.5) at $T$ = 5 K; (b) Temperature dependence of (0.5 0 0.5) magnetic peak intensity. Solid curves are Gaussian fits.

Figure 4. Plots of the comparison of observed and calculated intensities of various nuclear (a) and magnetic (b) diffraction peaks showing the goodness of the data refinement. Red lines are guides to eyes

Figure 1.

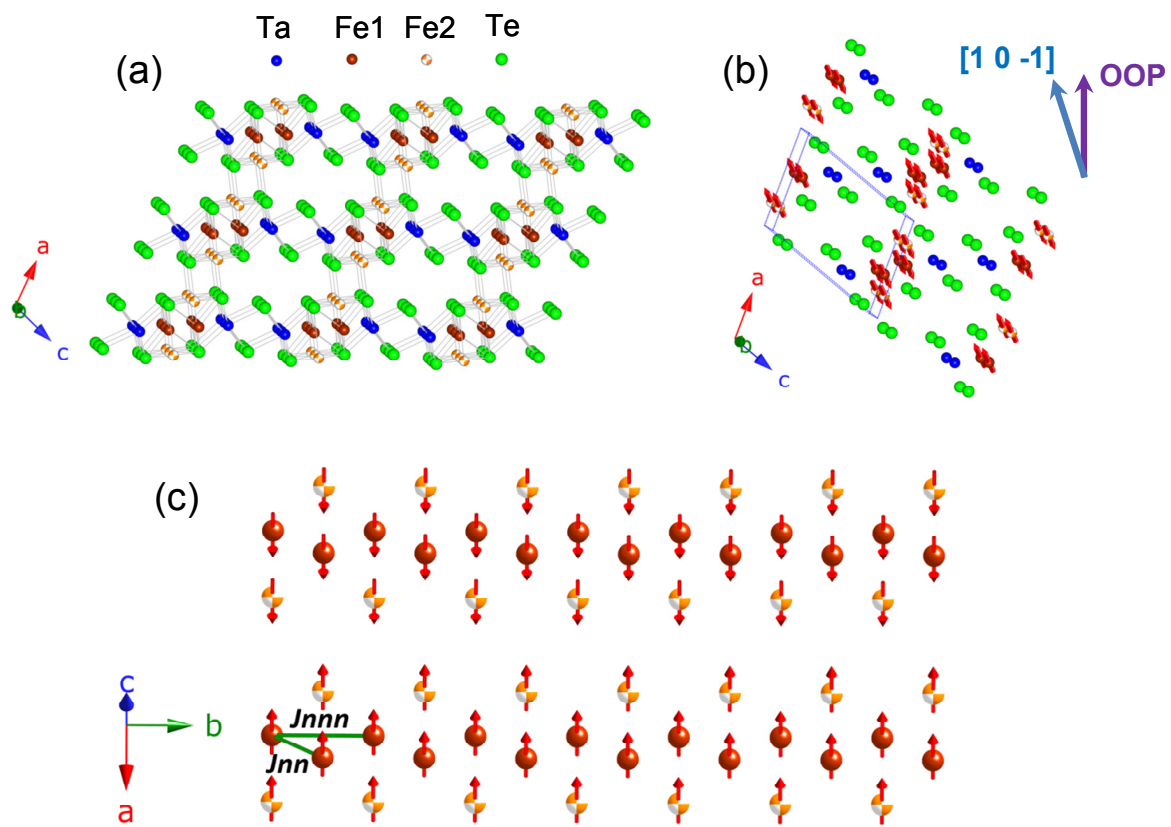

Figure 2.

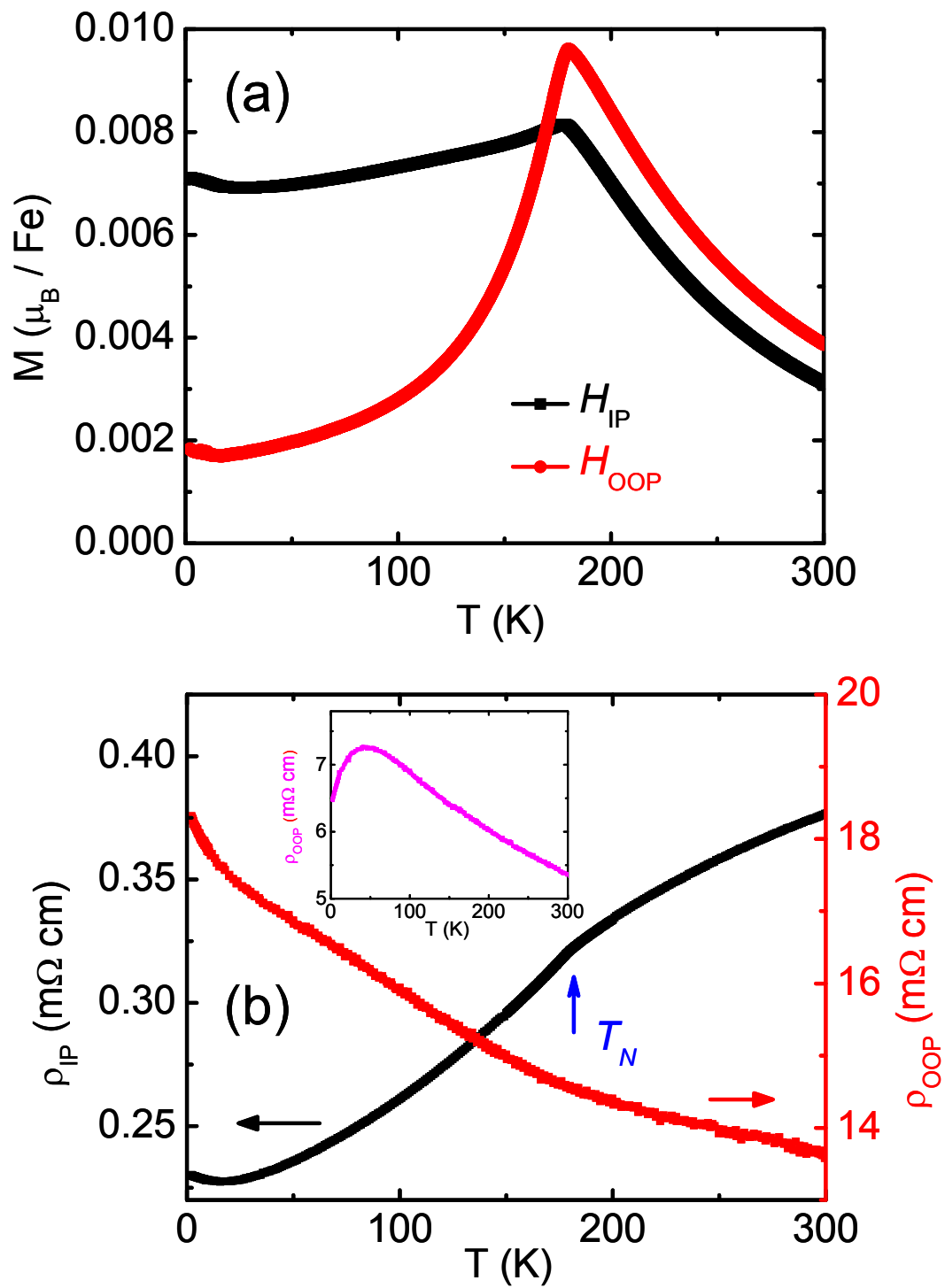

Figure 3.

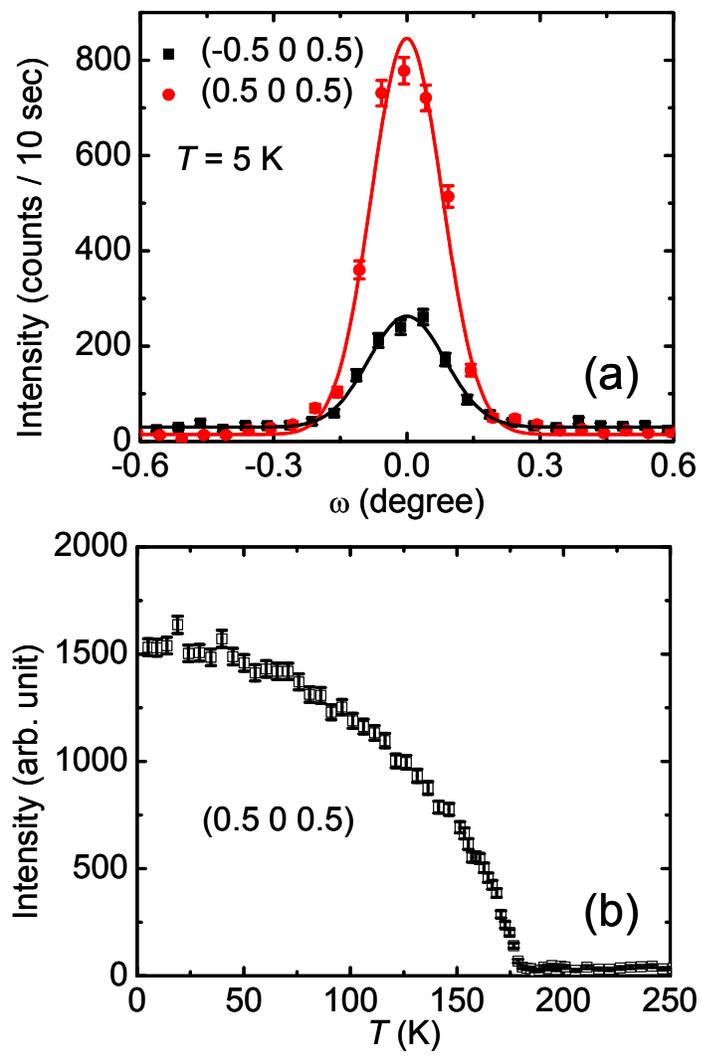

Figure 4.

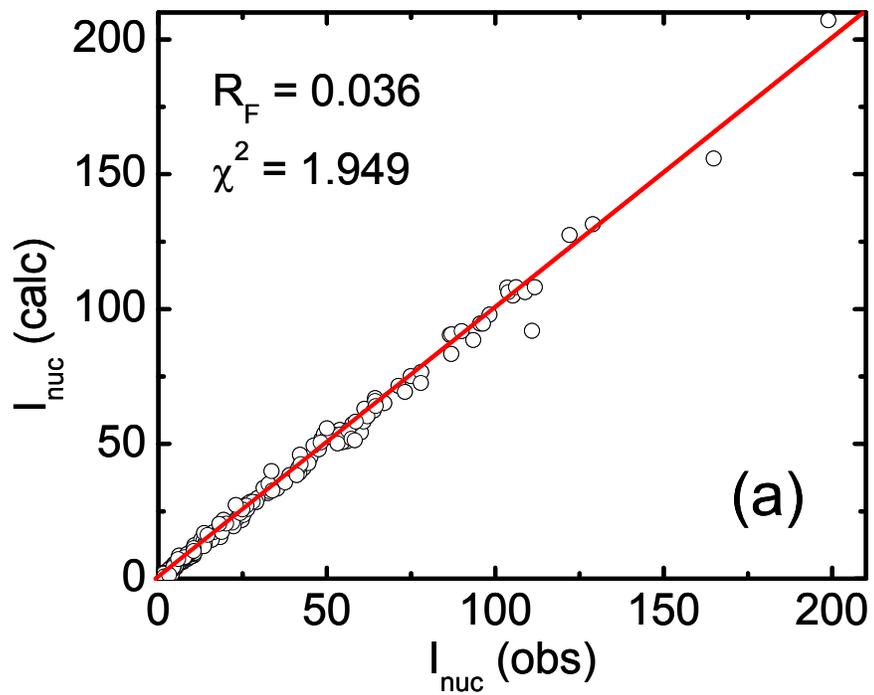

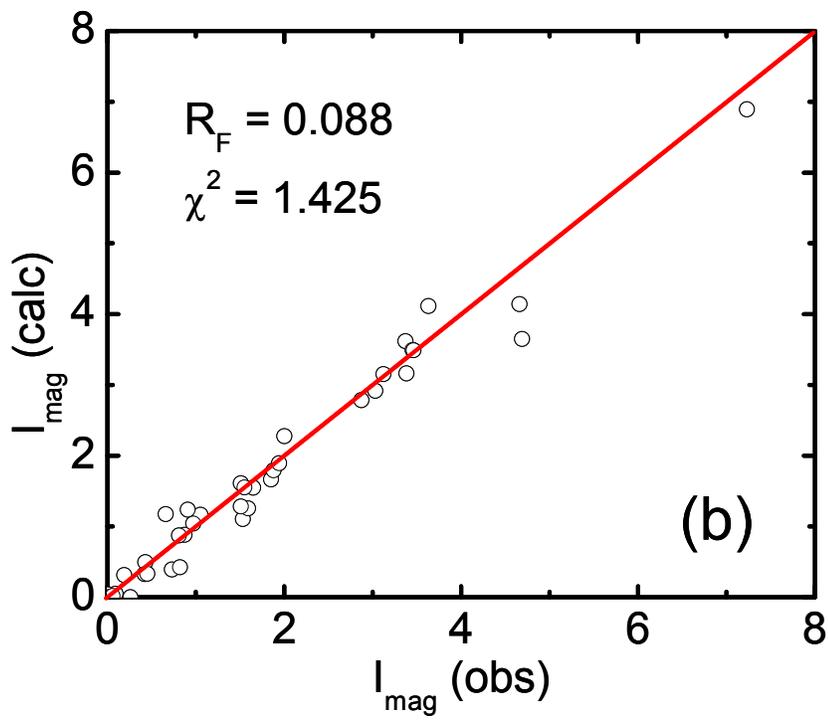